\documentclass[aps,showpacs,twocolumn,floatfix]{revtex4}
\usepackage{graphicx,times,amsmath,amssymb,epsfig}
\newcommand{\tr}{\mathop{\mathrm{Tr}}\nolimits}

\newcommand{\sech}{\mathrm{sech}} 
\newcommand{\nbar}{\bar{n}} 
 \def\la{\langle} \def\ra{\rangle}
\newcommand{\bw}{\begin{widetext}}
\newcommand{\ew}{\end{widetext}}
\newcommand{\sh}{\,\mathrm{sh}\,}
\newcommand{\ch}{\,\mathrm{ch}\,}
\def\nn{\nonumber} 
\begin{document}
\title{Squeezed vacuum as a universal quantum probe}
\author{Roberto Gaiba}
\affiliation{Dipartimento di Fisica dell'Universit\`a di Milano, I-20133 Milano, Italia.}
\author{Matteo G A Paris}
\affiliation{Dipartimento di Fisica dell'Universit\`a di Milano, I-20133 Milano, Italia.}
\affiliation{CNISM, Udr Milano Universit\`a, 20133, Milano, Italia.}
\affiliation{Institute for Scientific Interchange Foundation, I-10133 Torino, Italy}
\date{\today}
%%%%%
\begin{abstract}
We address local quantum estimation of bilinear Hamiltonians probed by
Gaussian states. We evaluate the relevant quantum Fisher information
(QFI) and derive the ultimate bound on precision.  Upon maximizing the
QFI we found that single- and two-mode squeezed vacuum represent an
optimal and universal class of probe states, achieving the so-called
Heisenberg limit to precision in terms of the overall energy of the
probe.  We explicitly obtain the optimal observable based on the
symmetric logarithmic derivative and also found that homodyne detection
assisted by Bayesian analysis may achieve estimation of squeezing with
near-optimal sensitivity in any working regime.  Besides, by comparison
of our results with those coming from global optimization of the
measurement we found that Gaussian states are effective resources, which
allow to achieve the ultimate bound on precision imposed by quantum
mechanics using measurement schemes feasible with current technology.
\end{abstract}
\maketitle
%%%%%%%%%%%%%%%%%%%%
\section{Introduction}
In this paper we address quantum estimation of unitary operations
for continuous variable systems. In particular we analyze the
estimation of the interaction parameter $\theta$ for unitaries of
the form $U_\theta = \exp\{-i \theta G\}$ where $G$ is a linear
or bilinear bosonic Hamiltonian of the form $G=a^\dag b + a
b^\dag$, $G=a^\dag b^\dag + a b$, or $G=a^{\dag 2} + a^2$,
$[a,a^\dag]=1$ and $[b,b^\dag]=1$ being mode operators. We are
interested in evaluating the ultimate bound on precision
(sensitivity), {\em i.e} the smallest value of the parameter that
can be discriminated, and to determine the optimal measurement
achieving those bounds.
\par
As a matter of fact, linear and bilinear interactions for bosonic
systems are a key ingredient for the development of continuous
variable quantum information processing \cite{r1,r2,r3,r4}. They are
usually realized by means of parametric processes, as single- and
two-mode squeezing, or by linear optical elements such as
phase-shifting and two-mode mixing. The precise characterization
of linear optical gates is also of interest in interferometry
\cite{par95,sand,camp}, absorption measurement \cite{per01} and
characterization of detectors \cite{ien}.
\par
\begin{figure}
\includegraphics[width=0.45\textwidth]{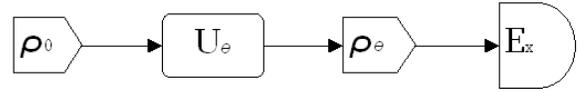}
\caption{General scheme for the indirect estimation of the
the unitary $U_\theta$ probed by the signal $\varrho_0$.
\label{f:sch}}
\end{figure}

In general, interaction parameters cannot be directly accessed
experimentally, and the estimation process consists in {\em
probing} the interaction by a known quantum signal $\varrho_0$,
which is measured after the interaction (see Fig. \ref{f:sch}).
The relevant constraint in the optimization of those schemes
concerns the total energy of the probe, which should be kept as
low as possible to avoid any possible modification or degradation
of the gate itself. Overall, the problem we are facing is that of
devising the optimal measurement, {\em i.e.} a positive
operator-valued measure (POVM) $\{E_x\}_{x\in{\cal X}}$, to be
performed on the probe $\varrho_\theta = U_\theta\varrho_0
U^\dag_\theta$ after the interaction, at fixed energy
$N=\hbox{Tr}[\varrho_0\,\sum_j n_j]$ of the incoming signal,
$\sum_j n_j$ being the total number operator of the involved
modes.
\par
The above problem may be properly addressed in the framework of
quantum estimation theory (QET) \cite{QET1,QET2,QET3}, which provides
analytical tools to find the optimal measurement according to some
given criterion.  In turn, there are two main paradigms in QET:
Global QET looks for the POVM minimizing a suitable cost
functional, averaged over all possible values of the parameter to
be estimated.  The result of a global optimization is thus a single
POVM, independent on the value of the parameter. On the other
hand, {\em local} QET looks for the POVM maximizing the Fisher
information, thus minimizing the variance of the estimator, at a
fixed value of the parameter \cite{hel67,BC94}.  Roughly speaking,
one may expect local QET to provide better performances since the
optimization concerns a specific value of the parameter, with some
adaptive or feedback mechanism assuring the achievability of the
ultimate bound \cite{gill00}.  Global QET has been mostly applied
to find optimal measurements and to evaluate lower bounds on
precision for the estimation of parameters imposed by unitary
transformations. For bosonic systems these include single-mode
phase \cite{Hol79,Dar98}, displacement \cite{Hel74}, squeezing
\cite{Mil94,Chi06} as well as two-mode transformations, e.g.
bilinear coupling \cite{per01}. Local QET has been applied to the
estimation of quantum phase \cite{Mon06} and to estimation
problems with open quantum systems and non unitary processes
\cite{Sar06}: to finite dimensional systems \cite{Hot06}, to
optimally estimate the noise parameter of depolarizing
\cite{Fuj01} or amplitude-damping \cite{Zhe06}, and for continuous
variable systems to estimate the loss parameter of a quantum
channel \cite{Mon07}.
\par
In this paper we consider the estimation the interaction parameters of
bilinear bosonic Hamiltonians from the perspective of local QET. In
particular, we focus our attention to measurement schemes as in Fig.
\ref{f:sch} with the probe state chosen within the set of Gaussian
states \cite{mars,r1,r2,r3,rg4}, which represents a class of signals
achievable with current technology.  We evaluate the relevant quantum
Fisher information (QFI) and derive the ultimate bound on precision.
Upon maximizing the QFI we found that single- and two-mode squeezed
vacuum represents an optimal and universal class of probe states,
achieving the so-called Heisenberg limit to precision in terms of the
overall energy of the probe. Remarkably, by comparison with results
coming from global optimization of the measurement
\cite{per01,Mil94,Chi06} we found that Gaussian states are effective
resources, which allow to achieve the ultimate bound on precision.
Besides, we found that homodyne detection assisted by Bayesian analysis
may achieve near-optimal sensitivity in any working regime.
\par
The paper is structured as follows: in the next Section we briefly
review local quantum estimation theory with some remarks on the
implementation of the optimal measurements. In Section
\ref{s:bulk} we evaluate the optimal measurements and the
corresponding bounds on precision for the local estimation of
bilinear couplings using Gaussian probes. In Section \ref{s:bayes}
we address estimation of squeezing using homodyne detection and
Bayesian analysis and show that near-optimal precision may be
achieved in any working regime.  In Section \ref{s:out} we compare
our results with those coming from global estimation and close the
paper with some concluding remarks.
%%%%%%%%%%%%%%%%%%%%
\section{Local quantum estimation theory}
In this section we review some concepts of local quantum estimation
theory \cite{LQE1,LQE2} which will be used in the rest of the paper.  
As a matter of fact, many quantities of interest in different
branches of physics cannot be directly accessed experimentally,
either in principle, as in the case of field measurement
\cite{Mab01}, or due to experimental impediments. In these cases,
one has to indirectly estimate the value of those physical
parameters by measuring a different observable, somehow related to
the quantity of interest. This indirect procedure of parameter
estimation implies an additional uncertainty for the measured
value, that cannot be avoided even in optimal conditions. The aim
of quantum estimation theory is to optimize the inference
procedure by minimizing this additional uncertainty.  In the
classical theory of parameter estimation the Cram\'{e}r-Rao Bound
\cite{Cra46} establishes a lower bound for the variance of any
unbiased estimator $\hat\theta$ of the parameter $\theta$. This
lower bound is given by the inverse of the so-called Fisher
Information (FI):
\begin{align}\label{cramer-rao}
\Delta\theta^2 \ge \frac{1}{F(\theta)}
\end{align}
where the Fisher Information is defined as
\begin{align}
F(\theta)
& = \sum_x p(x|\theta)
\left(\frac{\partial\ln p(x|\theta)}
{\partial\theta}\right)^2 \\
\label{FI}
\end{align}
Here $\theta$ is the parameter to be estimated, and $x$ denotes the
outcome of the measurement of the quantity $X$ related to $\theta$.
The notation $p(x|\theta)$ indicates the conditional probability of
obtaining the value $x$ when the parameter has the value $\theta$.
\par
A quantum analogue to Eq. (\ref{FI}) may be found starting from the Born rule
\begin{align}
p(x|\theta) = \mathrm{tr}[E_x \rho_\theta]
\end{align}
where $E_x$ are the elements of a positive operator-valued measure (POVM) and
$\rho_\theta$ is the density operator, parametrized by the quantity of
interest, describing the quantum state of the measured system.
The Fisher Information is then rewritten as
\begin{align}\label{FI2}
F(\theta) =  \sum_x \frac{\hbox{Re}\tr [\rho_\theta
E_x \Lambda_\theta]^2}{\tr [E_x \rho_\theta]}
\end{align}
where we introduced the Symmetric Logarithmic Derivative (SLD)
$\Lambda_\theta$, which is the self-adjoint operator defined as
\begin{align}\label{lambdaDEF}
\frac{\Lambda_\theta\rho_\theta+\rho_\theta\Lambda_\theta}{2}
\equiv \frac{\partial\rho_\theta}{\partial\theta}
\end{align}
It can then be shown \cite{hel67,BC94} that the Fisher Information (\ref{FI2}) is upper
bounded by the so-called \emph{Quantum Fisher Information} (QFI):
\begin{align}\label{ineqQFIdef}
F \le H \equiv \tr [\rho_\theta\Lambda_\theta^2]
\end{align}
In turn, the quantity $1/H$ represents an ultimate lower bound on
precision for any quantum measurement (followed by any classical
data processing) aimed to estimate the parameter $\theta$. The SLD
is itself an optimal measurement, that is, using the POVM $E_x$
obtained from the projectors over the eigenbasis of
$\Lambda_\theta$ we saturate the inequality (\ref{ineqQFIdef}).
\par
In this work we will focus on systems where the dependence of
$\rho_\theta$ from the parameter $\theta$ is generated by a family of
unitary transformations:
$\rho_\theta = U_\theta \rho_0 U^\dag_\theta$
where $U_\theta = \exp(-i\theta G)$, $G$ is the Hamiltonian that
generates the transformation and $\varrho_0$ is a given quantum
state used to probe the Hamiltonian process. In this case it is
possible to obtain an explicit formula for the SLD operator and
the QFI. At first we take the eigenbasis of $\rho_0$:
$\rho_0 = \sum_k p_k |\psi_k\rangle \langle \psi_k|$
From (\ref{lambdaDEF}) we can rewrite $\Lambda_\theta$ in this
basis as follows
\begin{align}\label{lambdaDEF2}
\Lambda_\theta=2i\sum_{jk}G_{jk}\frac{p_j-p_k}{p_j+p_k}
U_\theta|\psi_j\rangle\langle\psi_k|U^\dag_\theta
\end{align}
where $G_{jk}=\langle\psi_j|G|\psi_k\rangle$ are the matrix
elements of the generator $G$.
Eq.(\ref{lambdaDEF2}) shows that $\Lambda_\theta$ depends on $\theta$
only through the unitary transformation $U_\theta$.
As a consequence it is possible to define the operator $\Lambda_0$,
independent from $\theta$, such that $\Lambda_\theta=U_\theta
\Lambda_0 U^\dag_\theta$. It also follows that the quantum Fisher
information is independent from $\theta$. In fact,
$H  = \mathrm{Tr}[\rho_\theta \Lambda_\theta^2]
  = \mathrm{Tr}[U_\theta \rho_0 U^\dag_\theta
  U_\theta \Lambda_0^2 U^\dag_\theta]
   = \mathrm{Tr}[\rho_0 \Lambda_0^2]$.
Explicit formulas to calculate $H$ may be given in the eigenbasis 
of $\rho_0$
%\footnote{The sum over $n$ has to be taken
%only for those states $|\psi_n\rangle$ that have a nonzero probability
%$p_n$, while the sum over $k$ spans all the states of the basis of $\rho$.}:
\begin{align}
H 
%& =  4\sum_{nk}p_n \frac{(p_n-p_k)^2}{(p_n+p_k)^2}\, G_{nk}^2 
\label{Hmix1} 
%& =  2\sum_{nk}\frac{(p_n-p_k)^2}{p_n+p_k}\, G_{nk}^2  \\
& =  4\sum_{nk}p_n \frac{p_n-p_k}{p_n+p_k}\, G_{nk}^2  \\
& =  4\langle G^2 \rangle - 8\sum_{nk}\frac{p_k p_n}{p_n+p_k}\,
G_{nk}G_{kn} \label{Hmix2}
\end{align}
As we will see in the following, situations with a probe described
by a pure state $\rho_0=|\psi_0\rangle\langle\psi_0|$ are of particular
interest. In those cases the QFI reduces to the variance of the
generating Hamiltonian $G$, {\em i.e.} $H = 4\Delta G^2$.
In addition, for a pure state we have $\rho_\theta^2=\rho_\theta$ 
and thus $\Lambda_0 = 2i[\rho_0,G]$ {\em i.e.} 
\begin{align}
\Lambda_0 = 2i\sum_k \Big(G_{0k}|\psi_0\ra\la
\psi_k|-G_{k0}|\psi_k\ra\la\psi_0|\Big)
\label{SLDpure}\:.
\end{align}
%%%%%%%%%%%%%%%%
\section{Estimation of bilinear couplings}
\label{s:bulk}
In this Section we address the case of local estimation of various bilinear couplings
(single- and two-mode squeezing, two-mode mixing)
using Gaussian probes at fixed energy.
\subsection{Single-mode squeezing}
Here we consider the estimation of the parameter $\theta$ imposed by
the unitary transformation $\exp\left(-i\theta G\right)$,
where $G$ is the generating Hamiltonian
\begin{align}
G = \frac12 (a^\dag\,^2+a^2)
\end{align}
We analyze the precision achievable in the estimation of $\theta$
by using different classes of (Gaussian) probe states.
The measurement aimed to estimate $\theta$ is made on the
transformed state
\begin{align}
\rho_\theta = \exp\left(-i\theta G\right) \rho_0 \exp\left(-i\theta G\right)
\end{align}
At first we analyze the case of a Gaussian pure probe {\em i.e}
a squeezed coherent state of the form $\rho_0 = |\psi_0\rangle\langle\psi_0 |$
with $|\psi_0\rangle = S(r)D(\alpha)|0\rangle$, where
\begin{align}
D(\alpha) & = \exp\left[\alpha a^\dag - \alpha^* a\right] \\
S(r) & = \exp\left[\frac{r}{2}\left(a^\dag\,^2-a^2\right)\right]
\end{align}
and where, without loss of generality, we have chosen a real squeezing parameter $r$ and a
complex displacement $\alpha=xe^{i\phi}$.
Since $\rho_0$ is a pure state, the QFI will be given by
\begin{align}
H & =  4\Delta G^2 = \langle \left(a^\dag\,^2+a^2\right)^2
\rangle - \langle a^\dag\,^2+a^2 \rangle^2
%\\ & =  \la a^\dag\,^4 \ra + \la a^4 \ra + 2\la
%a^\dag\,^2 a^2 \ra + 4 \la a^\dag a \ra
%+2 -\la a^\dag\,^2 \ra^2 -\la a^2\ra^2 -2\la a^\dag\,^2\ra\la a^2\ra \nn
\end{align}
Upon evaluating  all the expectation values
we obtain:
\begin{align}
\Delta G^2 & =  - x^2\cos 2\phi \sinh 2r + (2N+1)\sinh^2 r + N + \frac{1}{2}
\end{align}
where  $N \equiv \la a^\dag a \ra=x^2+\sinh^2 r$ denotes the overall energy
of the probe signal. The signal optimization corresponds to the maximization
of $H$ over the state parameter with the constraint of fixed $N$.
The phase $\phi$ is a free parameter since it does not influence the total
energy. The choice $\cos 2\phi=-1$ maximizes $H$ leading to
\begin{align}
H=  4(N-\sinh^2 r)\sinh 2r + 4(2N+1)\sinh^2 r + 4N + 2
\end{align}
which grows monotonically with $\sinh^2 r$ and achieve its maximum
\begin{align}\label{Hmax1mode}
H_{\mathrm{max}}=8N^2+8N+2
\end{align}
for $\sinh^2 r=N$ and $\alpha=0$,
corresponding to a squeezed vacuum probe. Thus, to obtain the
maximum accuracy in the estimation of $\theta$ it is more
efficient to use all the energy in squeezing rather than field
amplitude.
\par
In order to see the effects of mixing we have also considered a class of
probes made by squeezed thermal states
\begin{align}
\rho_0 = \frac{1}{\bar{n}+1}\sum_k
\left(\frac{\bar{n}}{\bar{n}+1}\right)^k S(z)|k\rangle\langle k|S^\dag(z)
\end{align}
where the squeezing $z=re^{i\phi}$ is a complex number.
%This means that $\rho_0$ has the diagonal form
%\begin{align}
%\rho_0=\sum_k p_k |\psi_k\rangle\langle\psi_k|
%\end{align}
%where
%\begin{align}
%p_k & = & \frac{1}{\bar{n}+1}\left(\frac{\bar{n}}{\bar{n}+1}\right)^k \\
%|\psi_k\rangle & = & S(z)|k\rangle
%\end{align}
%and $z=re^{i\phi}$ is a complex number.
We are now dealing with a mixed state; the corresponding QFI is thus given by (\ref{Hmix1}).
The state vectors of the diagonal basis of $\rho_0$ and their associated probabilities are
\begin{align}
|\psi_k\rangle & = S(i\theta)|k\rangle \\
p_k & = \frac{1}{\bar{n}+1}\left(\frac{\bar{n}}{\bar{n}+1}\right)^k \label{thermalcoeff}
\end{align}
The matrix elements of the generator $G$ are
\begin{align}
G_{jk}  \equiv  & \langle k|S^\dag(z)\,\frac{a^\dag\,^2+a^2}{2}\,S(z)|k\rangle
\nn \\
 = &\,\frac{1}{2}\Big[\sqrt{(j+1)(j+2)}(\mu^2+\nu^*\,^2)\:\delta_{j+2,k}
\nn \\    & +\sqrt{(k+1)(k+2)}(\mu^2+\nu^2)\:\delta_{j,k+2} \nn \\
    & +(2k+1)\mu(\nu+\nu^*)\:\delta_{j,k}\Big]
\end{align}
where $\mu = \cosh r$, $\nu = e^{i\phi}\sinh r$.
From this and (\ref{Hmix1}) we get
\begin{align}
H = & 2\left(\cosh^4 r+\sinh^4 r+2\cos 2\phi \, \sinh^2 r \, \cosh^2 r\right)
\nn \\ & \times \frac{4\bar{n}^2+4\bar{n}+1}{2\bar{n}^2+2\bar{n}+1}
\end{align}
The energy constraint is now given by
\begin{align}
N = \bar{n}+(2\bar{n}+1)\sinh^2r
\end{align}
Maximization over the free parameter $\phi$ leads to $\phi=0$ and in turn to
\begin{align}
H = 2\frac{(4\nbar^2+4\nbar+1)(4N^2+4N+1)}{(2\nbar^2+2\nbar+1)(2\nbar+1)^2}
\end{align}
The maximum of this function is found when $\nbar=0$: again we are led to squeezed vacuum.
\par
As we have already discussed, the optimal measurement, {\em i.e.}
when the Fisher Information is equal to the QFI, is realized by
the SLD $\Lambda$. For squeezed vacuum probes we may use Eq.
(\ref{SLDpure}) and obtain
\begin{align}
\Lambda_0 = i\sqrt{2}(2N+1)S(r)\Big\{|0\ra\la 2|-|2\ra\la 0|\Big\}S^\dag(r)
\end{align}
Summarizing, the most convenient way of estimating a squeezing
parameter is to probe the transformation by a squeezed vacuum
probe. The corresponding QFI scales as $H \simeq 8 N^2$ in terms
of the overall energy of the probe.
%%%
\subsection{Two-mode mixing}
Here we consider the case where the generator $G$ is the two-mode
mixing Hamiltonian:
\begin{align}\label{generator2sq}
G = a^\dag b +a b^\dag
\end{align}
Let us first consider a probe state made by factorized squeezed
thermal states:
\begin{align}
\label{rho0disentangled}
\rho_0 = [S_a(r) \otimes S_b(s)] \nu_a \otimes \nu_b [S_a^\dag(r) \otimes S_b^\dag(s)]
\end{align}
where $\nu_{a,b}$ are the density matrices of thermal states:
\begin{align}
\label{nuthermal}
\nu_k = \frac{1}{(\bar{n}_k+1)}
\sum_{n} \left(\frac{\bar{n}_k}{\bar{n}_k+1}\right)^n |n\rangle\langle n|
\end{align}
For a two-mode system the formula (\ref{Hmix1}) for the Quantum Fisher Information becomes
\begin{align}\label{H2modes}
H = 4\sum_{jkmn}p_{jk} \frac{p_{jk}-p_{mn}}{p_{jk}+p_{mn}} G_{jkmn}G_{mnjk}
\end{align}
where $p_{kn} = p_k p_n$, the thermal coefficients (\ref{thermalcoeff}).
The Heisenberg evolution of the mode operators
\bw\begin{align}
S_a^\dag(r)S_b^\dag(s)\left(a^\dag b +a b^\dag\right)S_a(r)S_b(s)
& = \cosh(r+s)(a^\dag b + ab^\dag)+\sinh(r+s)(ab+a^\dag b^\dag)
\end{align}
allows to calculate the matrix elements of $G$
\begin{align}
G_{jkmn} = & \langle j,k| S_a^\dag(r)S_b^\dag(s)\left(a^\dag b +
      a b^\dag\right)S_a(r)S_b(s)|m,n\rangle \nn \\
= & \cosh(r+s) \left(\sqrt{(m+1)(k+1)}\delta_{j=m+1}\delta_{n=k+1}
   +\sqrt{(j+1)(n+1)}\delta_{m=j+1}\delta_{k=n+1}\right) \nn \\
  & +\sinh(r+s)\left(\sqrt{(j+1)(k+1)}\delta_{m=j+1}\delta_{n=k+1}
  +\sqrt{(m+1)(n+1)}\delta_{j=m+1}\delta_{k=n+1}\right)
\end{align}
The resulting QFI reads as follows
\begin{align}\label{QFI2modes}
H & = 4\left[\sinh^2(r+s)\left(\frac{(\nbar_a
-\nbar_b)^2}{2\nbar_a\nbar_b+\nbar_a+\nbar_b}
+\frac{(\nbar_a+\nbar_b+1)^2}{2\nbar_a\nbar_b
+\nbar_a+\nbar_b+1}\right)
+\frac{(\nbar_a-\nbar_b)^2}{2\nbar_a\nbar_b
+\nbar_a+\nbar_b}\right]
\end{align}\ew
The total photon number of the system is given by the sum
\begin{align}\label{N2modes}
N = \nbar_a+\nbar_b+(2\nbar_a+1)\sinh^2r+(2\nbar_b+1)\sinh^2s
\end{align}
The QFI (\ref{QFI2modes}) has no point of gradient zero that is
compatible with the energy bound (\ref{N2modes}).  Since it is a
continuous function, to find its maximum we need to investigate its
value at the borders of its domain.
Let us first consider the case $\nbar_a=\nbar_b=0$,
i.e. a probe made by two disentangled squeezed vacuums.
The energy and the QFI become respectively
\begin{align}
N & = \sinh^2r+\sinh^2s \\
H & = 4\sinh^2(r+s)
\end{align}
The maximum of this function is reached when $r=s$, which gives
\begin{align}
H_1=4N^2+8N
\end{align}
The second possible case is given by two thermal states, when $r=s=0$.
The QFI becomes
\begin{align}
H & = \frac{4(N-2\nbar_b)^2}{N+2(N-\nbar_b)\nbar_b}
\end{align}
whose maximum is
\begin{align}
H_2=4N & \;\;\; \mathrm{when} \;\;\; \nbar_a=0 \;\;\; \mathrm{or}
\;\;\; \nbar_b=0
\end{align}
i.e. when one of the states is at zero temperature.
The last possible combination is given by a thermal state
and a squeezed vacuum, for $r=0$, $\nbar_b=0$.
Energy and QFI reduce to
\begin{align}
N & = \nbar_a+\sinh^2s \\
H & =
4\left[(2\nbar_a+1)\sinh^2s+\nbar_a\right]=4[N+2\nbar_a(N-\nbar_a)]
\end{align}
The optimal QFI is obtained when the energy is equally distributed
between the thermal state and the squeezed state,
$\nbar_a=\sinh^2s=\frac{N}{2}$: \begin{align} H_3=2N^2+4N \end{align}
Thus we see that the maximum Fisher information is obtained using
two equally squeezed vacuums. Since this is the combination of two
pure states, we can use (\ref{SLDpure}) to obtain the SLD that
realizes the optimal measurement:
\begin{align}
\Lambda_0 = 2i\sqrt{N(N+2)} S_a S_b\Big(|0,0\ra\la 1,1|-|1,1\ra\la 0,0|\Big)S_a^\dag S_b^\dag
\end{align}
In order to investigate the role of entanglement in the estimation
procedure we consider the probe prepared the state
\begin{align}\label{rho2entangled}
\rho_0 & = |\psi_{00}\rangle\langle\psi_{00}| \\
|\psi_{jk}\rangle & \equiv |\psi_{jk}(\phi,\lambda)\rangle\rangle = U(\phi)T(\lambda) |j,k\ra
\end{align}
where
\begin{align}
U(\phi) & = \exp[-i\phi(ab^\dag+a^\dag b)] \\
T(\lambda) & = \exp[-i\lambda(ab+a^\dag b^\dag)]
\end{align}
The probe is transformed into $\rho_\theta = e^{-i\theta G}\rho_0 e^{i\theta G}$ where again
we are using the generator (\ref{generator2sq}).
Since we are dealing with a pure state, the QFI is
\begin{align}
H & = 4\Delta G^2 = 16 \cosh^2|\lambda|\sinh^2|\lambda|\left(1-4\cos^2\phi\sin^2\phi\right)
\end{align}
where we have used the Heisenberg evolution of the mode operators.
The energy constraint is given by
\begin{align}
N=\langle a^\dag a \rangle + \langle b^\dag b \rangle = 2\sinh^2|\lambda|
\end{align}
thus the QFI can be rewritten as
\begin{align}
H & = \left(4N^2+8N\right)\left(1-4\cos^2\phi \sin^2\phi\right)
\end{align}
Since $\partial_\phi N=0$, we can freely choose a value for
$\phi$, in order to maximize $H$. The maximum Fisher information
is obtained for $\cos 4\phi=1$ and corresponds to $H=4N^2+8N$,
{\em i.e.} no improvement is obtained using an entangled probe.
The SLD operator that realizes the optimal measurement is found
using (\ref{SLDpure}): \bw\begin{align} \Lambda_0 & =
2i\sqrt{2N(N+1)}
\Big\{|\psi_{00}\rangle\langle\psi_{20}|+|\psi_{00}\rangle\langle\psi_{02}|
-|\psi_{20}\rangle\langle\psi_{00}|-|\psi_{02}\rangle\langle\psi_{00}|\Big\}
\end{align}\ew
%%%
\subsection{Two-mode squeezing}
The procedure used for the case of two-mode mixing may be analogously applied
when the generator $G$ is given by the two-mode squeezing Hamiltonian:
\begin{align}\label{generator2mix}
G=ab+a^\dag b^\dag
\end{align}
First we analyze the case of an initial density matrix, see
(\ref{rho0disentangled}), that describes two disentangled squeezed
thermal states.  The same steps done to obtain (\ref{QFI2modes}) can be
repeated, using the Hamiltonian (\ref{generator2mix}) instead of
(\ref{generator2sq}).
The QFI for this particular case is thus given by
\bw\begin{align}
H & =
4\left[\sinh^2(r+s)\left(\frac{(\nbar_1-\nbar_2)^2}{2\nbar_1\nbar_2+\nbar_1+\nbar_2}
+\frac{(\nbar_1+\nbar_2+1)^2}{2\nbar_1\nbar_2+\nbar_1+\nbar_2+1}\right)
+\frac{(\nbar_1+\nbar_2+1)^2}{2\nbar_1\nbar_2+\nbar_1+\nbar_2+1}\right]
\end{align}\ew
The maximum of this function is once again obtained when $\nbar_1=\nbar_2=0$ and $r=s$, i.e. when
the probe is made by two equally squeezed vacuum states. This max is
\begin{align}
H_{\mathrm{max}} = 4(2N+1)^2
\end{align}
where $N=2\sinh r$.
The corresponding SLD reads as follows
\begin{align}
\Lambda_0 = 2i(N+1)S_a S_b\Big(|0,0\ra\la 1,1|-|1,1\ra\la 0,0|\Big)S_a^\dag S_b^\dag
\end{align}
The same can be done for the case of a probe such as (\ref{rho2entangled}).
The corresponding QFI is given by
\begin{align}
H = & 8\cosh^2|\lambda| \left[(\cos^2\phi-\sin^2\phi)^2\cos
(2\arg\lambda)\sinh^2|\lambda| \right. \nonumber \\ & \left. +2\sinh^2|\lambda|+1\right]
\end{align}
The maximum Fisher Information $H_{\mathrm{max}} = 4N^2+8N$ is achieved when
$\cos(\arg 2\lambda)=1$ and $\cos 2\phi=1$
and using the SLD
\begin{align}
\Lambda_0 = 2i(2N+1)\Big(|\psi_{00}\ra\la\psi_{11}|-|\psi_{11}\ra\la\psi_{00}|\Big)
\end{align}
%%%%
\section{Estimation of squeezing by homodyne detection}
\label{s:bayes}
In Section \ref{s:bulk} we have shown that squeezed vacuum is the
optimal reference Gaussian state to estimate the parameter of a
squeezing transformation. However, the optimal measurement
maximizing the QFI, that is the SLD, is not realizable with
current technology. It is thus of interest to investigate whether
a feasible measure may be used to effectively probe the perturbed
squeezed vacuum. We focus to the case of single-mode squeezing
estimation; an analogue analysis may be performed for two-mode
operations. Our approach is to exploit homodyne detection to
measure field-quadrature:
\begin{align}\label{quadratura}
x_\alpha & = \frac12 \left(ae^{-i\alpha}+a^\dag e^{i\alpha}\right)
\end{align}
and inferring the squeezing parameter through the results
obtained with multiple homodyne measurements. 
The homodyne probability $p(x|\theta)$ is given by
\begin{align}
p(x|\theta) & = \mathrm{Tr}[\rho_\theta\: \Pi_x (\theta)]
\end{align}
$\Pi_x=|x\ra_\theta{}_\theta\la x|$ being the spectral measure of the 
quadrature (\ref{quadratura}).
The resulting distribution  for a squeezed vacuum 
to which an unknown squeezing has been applied, 
is a zero mean ($\hbox{Tr}\left[\rho_\theta\,x_\alpha \right]= 0$)
Gaussian distribution 
\begin{align}\label{pxgauss}
p(x|\theta) & = \frac{1}{\sqrt{2\pi\Sigma^2_\theta}}
\exp\left\{-\frac{x^2}{2\Sigma^2_\theta}\right\}
\end{align}
with variance (see the Appendix for details on the derivation)
\begin{align}\label{sigma2theta}
\Sigma^2_\theta & = \cos(2\alpha)\sqrt{N(N+1)} \nonumber \\ 
& +\left(N+\frac12\right)[\cosh(2\theta)+\sin(2\alpha)\sinh(2\theta)]
\end{align}
The reason to choose homodyne detection 
is that the classical Fisher information (\ref{FI}) of the homodyne 
distribution $p_\alpha(x|\theta)$ may be optimized over $\alpha$ in
order to achieve the same scaling as the QFI versus the energy of the
probe. Being $\alpha_1 = \arg\max_\alpha F_\alpha(\theta)$ we have
\begin{align}
\cos\alpha_1 & = \left[-\sqrt{\frac12
-\frac{\sqrt{N(N+1)}}{(1+2N)\cosh\theta-\sinh\theta}}\right] \\ 
F_{\alpha_1}(\theta) & \stackrel{N\gg 1}{\simeq} 8 N^2  
\label{optFh}
\end{align}
This means that homodyne detection with optimized phase $\alpha$ is
a good candidate to achieve ultimate bounds to precision, as far as
it saturates the classical Cramer-Rao bound.
Indeed, Von Mises-Bernstein-Laplace theorem ensures that Bayesian 
\emph{a posteriori} distribution $p(\theta|\{x\}_M)$, representing the probability 
of the squeezing to be $\theta$ given the homodyne sample $\{x\}_M$, 
converges asymptotically to a Gaussian distribution, centered in the
true value with variance saturating the Cramer-Rao bound. In other
words, Bayesian estimators are asymptotically unbiased and efficient.
In the following, we thus discuss in some details estimation of
squeezing by homodyne detection and Bayesian  analysis.
We consider a large number $M$ of homodyne measurements on repeated 
preparations of the same
system.  Since the measurements are independent, the a posteriori 
distribution is proportional to the product of the single data
distribution
\begin{align}\label{pcond2}
p(\theta|\{x\}_M) & \propto
\prod_{k=1}^M p(\theta|x_k) = \prod_{k=1}^M \frac{p(x_k|\theta)p(\theta)}{p(x_k)}
\end{align}
where we repeatedly used the Bayes Theorem.
$p(\theta)$ is the \emph{a priori} distribution of the parameter,
$p(x)$ the overall probability of the outcome $x$,
while $p(x|\theta)$ is the probability to obtain the outcome $x$ when
the squeezing parameter is $\theta$.
The probability $p(\theta|\{x\}_M)$ has to be normalized, Eq.(\ref{pcond2})
thus rewrites as
\begin{align}\label{pcond3}
p(\theta|\{x\}_M) & = \frac1A\, p(\theta)^M \prod_{k=1}^M \frac{p(x_k|\theta)}{p(x_k)}
\end{align}
where $A$ is the normalization constant given by
\begin{align}
A & = \int_{-\infty}^{+\infty}p(\theta)^M \prod_{k=1}^M \frac{p(x_k|\theta)}{p(x_k)}
\end{align}
We assume to have no \emph{a priori} information on the squeezing
$\theta$ {\em i.e.} we take $p(\theta)$ as a uniform function. Notice
also that the product of the distributions $p(x_k)$ does not depend 
on $\theta$ and it cancels out due to normalization.  Finally, since we wish
to perform a large number $M\gg 1$ of measurements, the
product in (\ref{pcond3}) will contain many repeated elements:
each outcome $x$ is obtained a number of times
proportional to its probability $p(x|\theta^*)$, being $\theta^*$
the \emph{true} (and unknown) value of the squeezing parameter.
We can then re-order the product so that its index now runs through
all possible values of $x$:
\begin{align}\label{pcond1}
p(\theta|\{x\}_M) & \simeq \frac1A \prod_x p(x|\theta)^{M p(x|\theta^*)} \nn \\
& = \frac1A \exp\left\{M\int p(x|\theta^*)\ln p(x|\theta)dx\right\}
\end{align}
where we have taken a limit to the continuum for the variable $x$.
The integral in (\ref{pcond1}) can be solved leading to
\begin{align}
\int_{-\infty}^{+\infty}p(x|\theta^*)\ln p(x|\theta)dx
& = -\frac12 \left[\frac{\Sigma_*^2}{\Sigma^2_\theta}+\ln(2\pi\Sigma^2_\theta)\right]
\end{align}
where we have introduced the short notation $\Sigma^2_* 
\equiv \Sigma^2_{\theta^*}$. Overall, we obtain
\begin{align}\label{pcond4}
p(\theta|\{x\}_M) & =
\frac1A \left[\Sigma^2_\theta \exp\left(\frac{\Sigma_*^2}{\Sigma^2_\theta}\right)\right]^{-M/2}
\end{align}
where we have redefined $A$ so to include all terms independent from $\theta$.
The mean $\bar\theta$ of the \emph{a posteriori} distribution
$p(\theta|\{x\}_M)$ is our estimator and the variance
$\Delta\theta^2$ the corresponding confidence interval
\begin{align}
\bar{\theta}   & = \int_{-\infty}^{+\infty}\!\!\! d\theta\: \theta\: p(\theta|\{x\}_M)  \\
\Delta\theta^2 & = \int_{-\infty}^{+\infty}\!\!\! d\theta\:
(\theta-\bar{\theta})^2\: p(\theta|\{x\}_M)\:.
\end{align}
An optimal value for the homodyne phase $\alpha$ is obtained upon
minimizing the variance of the a posteriori distribution. Besides 
the value $\alpha_1$ reported above we found that optimal scaling 
($\propto M^{-1} N^{-2}$) of the variance may be achieved also for
the phase value $$\alpha_2 = - \hbox{sign}(\theta^*)\arccos \left[
\sqrt{\sech(2 \theta^*)\sinh^2 \theta^*}\right]\,,$$ which, remarkably, 
is independent on the probe energy $N$ (indeed, we have $\alpha_1 =
\alpha_2 + O(1/N)$).
\par
%%%%
\begin{figure}[h]
\includegraphics[width=0.22\textwidth]{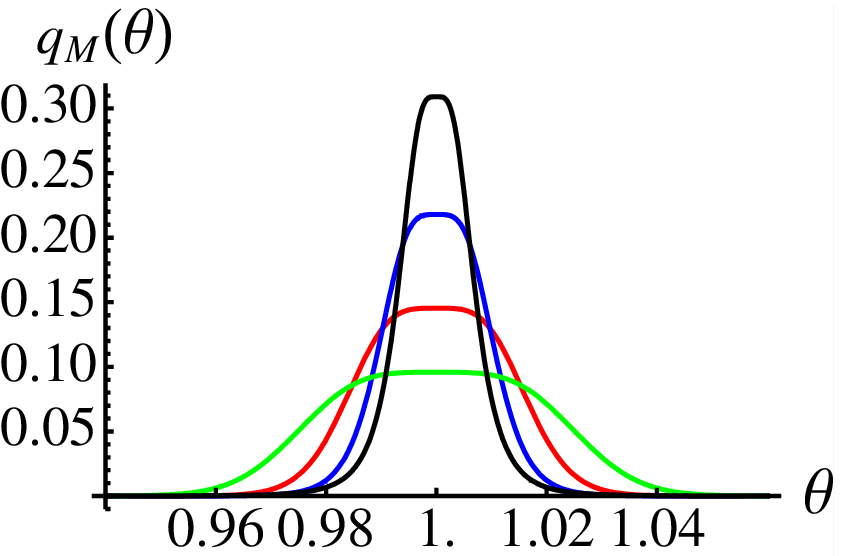}
\includegraphics[width=0.25\textwidth]{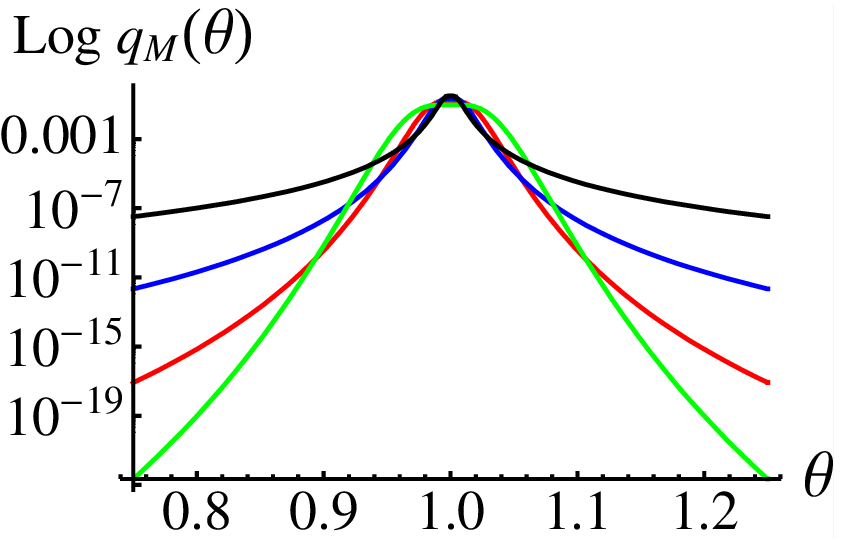}
\caption{(Color online) Left: Rescaled a posteriori distribution $q_M(\theta)$ for 
$M=5, N=40$ (black), $M=10, N=20$ (blue), $M=20, N=10$ (red), $M=40,
N=5$ (green). Right: LogPlot of the rescaled a posteriori
distribution. for the same values of the parameters.
\label{f:p}}
\end{figure}
%%%%
In Fig. \ref{f:p} we report the rescaled distribution
$q_M(\theta)= p(\theta|\{x\}_M)/(M N)$ for different values of the probe energy and 
the number of measurements, we also report $p(\theta|\{x\}_M)/(M N)$ in a
logarithmic scale to enlighten the differences in the distribution tails. 
As it is apparent from the plots the relevant parameter is the energy of the probe.
For highly excited probes, {\em i.e} for $N\gg 1$, we expand $\Sigma^2_\theta$ as
\begin{align}
\Sigma^2_\theta = & \left(N+\frac12\right)\,\left[\cos(2\alpha)+\cosh(2\theta)+
\sin(2\alpha)\sinh(2\theta)\right] \nn \\
& -\frac{\cos(2\alpha)}{8N}+O\left(\frac{1}{N^2}\right)
\end{align}
and neglect all orders scaling as $N^{-2}$ or higher.
Upon choosing the homodyne phase $\alpha_2$  we have
\begin{align}
\Sigma^2_* & \simeq \frac{\sech(2\theta^*)}{8N} \label{sigmapprox1} \\
\Sigma^2_\theta   & \simeq \sech(2\theta^*)
\left[(2N+1)\sinh^2(\theta-\theta^*)+\frac{1}{8N}\right] \label{sigmapprox2}
\end{align}
Upon substituting (\ref{sigmapprox1}) and (\ref{sigmapprox2}) into (\ref{pcond4}) we 
see explicitly that $p(\theta|\{x\}_M)=p(\theta-\theta^*|\{x\}_M)$ and that the 
estimator is indeed unbiased, {\em i.e} $\bar{\theta} = \theta^*$. We
also found that the variance is independent from
the true value of the squeezing $\theta^*$:
Numerical computation shows that
the variance $\Delta\theta^2$ scales
as $\displaystyle\sim\frac{1}{4 M N^2}$ for large $N$, that is, 
apart from a factor two, the same scaling of the inverse
of the QFI (\ref{Hmax1mode}).
Notice that the optimal phase $\alpha_2$,
depends on $\theta^*$, which is the unknown parameter that we 
are trying to estimate. This is consistent with the local nature 
of the estimator procedure. From a practical point of view
this means that some kind of feedback mechanism or adaptive
technique should be employed to adjust the phase of the
homodyne detector \cite{gill00,phfed}.
We conclude that homodyne detection with Bayesian analysis %estimator
is a robust and accurate estimation technique for the squeezing
parameter. Remarkably, this scheme may be implemented with current
technology.
%%%%%
\section{Conclusions}
\label{s:out}
In this paper we have addressed local quantum estimation of bilinear
Hamiltonians probed by Gaussian states. We evaluated the relevant
quantum Fisher information (QFI) thus obtaining the ultimate bound
on precision. Upon maximizing the QFI we found that single- and two-mode
squeezed vacuum represent an optimal and universal class of probe states,
achieving the so-called Heisenberg limit to precision in terms of the
overall energy of the probe. For two-mode operations no improvement
may be obtained using entangled probes.
\par
It is worth noting that the Heisenberg scaling $\Delta \theta \sim N^{-1}$
in terms of the overall energy of the probe may be achieved also using
global quantum estimation techniques (see {\em e.g.} \cite{per01} for
the case of two-mode mixing). In
that case, however, optimization of the probe have been performed over
the whole set of quantum states, not focusing on Gaussian states.
In  turn, this means that Gaussian states are effective resources, which
allow to achieve the ultimate bound on precision imposed by quantum
mechanics using measurement schemes feasible with current technology.
This has been confirmed by a Bayesian analysis applied to the estimation
of squeezing by  homodyne detection, which achieves near-optimal
sensitivity in any working regime, {\em i.e} for any (true) value of
the squeezing parameter. For the estimation of squeezing, Heisenberg
scaling for Gaussian probes has been also found exploiting global
strategies \cite{Chi06}. In that case, however, though the measurement
does not depend on the value of the parameter, there is a strong dependence
on the probe states.  We have also explicitly obtained the optimal
observables based on the symmetric logarithmic derivative, which however
do not correspond, in general, to a feasible detection scheme.
\par
We conclude that Gaussian states and Gaussian measurements assisted by
Bayesian analysis represent robust and accurate resources for the
estimation of unitary operations of interest in continuous variable
quantum information.
%%%
\bw
\section{Appendix}
Here we show how Eq.(\ref{sigma2theta}) is obtained. We start from the identity
\begin{align}
S^\dag(z) a S(z) & = \mu a + \nu a^\dag
\end{align}
where $\mu =  \cosh |z|$ and $\nu =  e^{i \arg z} \sinh |z|$.
In turn this leads to
\begin{align}
S^\dag(r)S^\dag(i\theta) a S(i\theta) S(r)
& = (a \cosh r + a^\dag \sinh r)\cosh\theta+i(a^\dag \cosh r
+ a \sinh r)\sinh\theta \nn \\ & = a(\cosh r \cosh\theta + i
\sinh r \sinh\theta)+a^\dag(\sinh r \cosh\theta+i\cosh r \sinh\theta)
\end{align}
and then
\begin{align}
S^\dag(r)S^\dag(i\theta) x_\alpha S(i\theta)S(r)
& = \frac12 \Big\{e^{i\alpha}\Big[a^\dag(\ch r \ch\theta-i\sh r \sh\theta)
    +a(\sh r \ch\theta-i\ch r \sh\theta)\Big]+h.c.\Big\} \nn \\
& = \frac12 \Big\{ a^\dag \Big[(\ch r \ch\theta-i\sh r \sh\theta)e^{i\alpha}
    +(\sh r \ch\theta+i\ch r \sh\theta)e^{-i\alpha}\Big]+h.c.\Big\}
\end{align}
When the square of this operator is averaged in the vacuum $\la 0|...|0\ra$,
only one of the four terms $a^2$, $a^\dag a$, $aa^\dag$ and $a^\dag\,^2$ survives, namely
$\la 0|aa^\dag|0\ra=1$.
The equation then simplifies to
\begin{align}
\Sigma^2_\theta = & \frac12 \Big\{e^{2i\alpha}(\ch r \ch\theta
-i \sh r \sh\theta)(\sh r \ch\theta-i \ch r \sh\theta)
+e^{-2i\alpha}(\sh r \ch\theta +i \ch r \sh\theta)(\ch r
\ch\theta +i \sh r \sh\theta) \nn \\
& + \, (\ch r \ch\theta -i \sh r \sh\theta)(\ch r \ch\theta
+i \sh r \sh\theta) +    (\sh r \ch\theta -i \ch r \sh\theta)
(\sh r \ch\theta +i \ch r \sh\theta)\Big\} \nn \\
= &  \frac12 \Big\{\sinh(2r)\cos(2\alpha)+\cosh(2r)
[\cosh(2\theta)+\sin(2\alpha)\sinh(2\theta)]\Big\} \nn \\
= &  \cos(2\alpha)\sqrt{N(N+1)}+\left(N+\frac12\right)
[\cosh(2\theta)+\sin(2\alpha)\sinh(2\theta)]
\end{align}
where we used $N=\sinh^2r$.
\ew
%%%%

%%%%%%%%%%%%%%%%%%%%%%%%%%%%%%%%%%%%%%%%%%%%%%%%%%%%%%%%%%%%


\begin{thebibliography}{99}
\bibitem{r1}
B. G. Englert, K.Wodkiewicz, Int. J. Quant. Inf. {\bf 1},
153 (2003).
\bibitem{r2}
A. Ferraro, S. Olivares, M.  G. A. Paris {\em
Gaussian states in quantum information}, (Bibliopolis, Napoli,
2005).
\bibitem{r3}
F. Dell'Anno, S. De Siena, F. Illuminati Phys. Rep. {\bf 428}, 53
(2006).
\bibitem{r4}
{\em Quantum information with continuous variables of atoms and light}
N. J. Cerf, G. Leuchs, E. S. Polzik (Eds.), (Imperial College Press,
London, 2007) and reference therein. 
\bibitem{rg4} X.-B. Wang et al., Phys. Rep. 448, 1 (2007).
\bibitem{par95} M. G. A. Paris, Phys. Lett A {\bf 201}, 132 (1995).
\bibitem{sand} B. C. Sanders, G .J. Milburn,
Phys. Rev. Lett. {\bf 75}, 2944 (1995).
\bibitem{camp} R. A. Campos {\it et al}, Phys. Rev. A {\bf 40}, 1371 (1989).
\bibitem{per01} G. M. D'Ariano, M. G. A. Paris, P. Perinotti,
J. Opt. B  {\bf 3}, 337 (2001).
\bibitem{ien} G. Brida, M.Genovese, M. Gramegna, Las. Phys. Lett. {\bf
3}, 115 (2006).
\bibitem{QET1}
S. D. Personick, IEEE Trans. Inf. Th. {\bf 17}, 240 (1971).
\bibitem{QET2}
C.  W. Helstrom, {\em Quantum detection and estimation theory},
(Academic Press, NY, 1976);
\bibitem{QET3}
A. S. Holevo, {\em Probabilistic and statistical aspects of quantum theory},
(North-Holland, Amst'm, 1982); A.S. Holevo, {\em Statistical Structure of Quantum Theory},
Lect. Not Phys. {\bf 61} (Springer, Berlin, 2001).
\bibitem{hel67} C. W. Helstrom, Phys. Lett. A {\bf 25}, 1012 (1967).
\bibitem{BC94} S. L. Braunstein, C. M. Caves, Phys. Rev. Lett. {\bf 72}
3439 (1994); S. L. Braunstein, C. M. Caves, G. J. Milburn, Ann. Phys.
{\bf 247}, 135 (1996).
\bibitem{gill00} O. E. Barndorff-Nielsen, R. D. Gill, J. Phys. A {\bf
33}, 4481 (2000).
\bibitem{Hol79} A. S. Holevo, Rep. Math. Phys. {\bf 16}, 385 (1979).
\bibitem{Dar98} M. D'Ariano et al., Phys. Lett. A {\bf 248}, 103 (1998).
\bibitem{Hel74} C. W. Helstrom, Found. Phys. {\bf 4}, 453 (1974).
\bibitem{Mil94} G. J. Milburn et al., Phys. Rev. A {\bf 50}, 801 (1994).
\bibitem{Chi06} G. Chiribella et al., Phys. Rev. A {\bf 73}, 062103 (2006).
\bibitem{LQE1} D. C. Brody, L. P. Hughston, Proc. Roy. Soc. Lond. A
{\bf 454}, 2445 (1998); A {\bf 455}, 1683 (1999). 
\bibitem{LQE2} S. Amari and H. Nagaoka, {\em Methods of Information
Geometry},  Trans. Math. Mon. {\bf 191}, AMS (2000).
\bibitem{Mon06} A. Monras, Phys. Rev. A {\bf 73}, 033821 (2006).
\bibitem{Sar06} M. Sarovar, G. J. Milburn, J. Phys. A {\bf 39}, 8487
(2006).
\bibitem{Hot06} M. Hotta et al., Phys. Rev. A {\bf 72}, 052334 (2005);
J. Phys. A {\bf 39} (2006).
\bibitem{Fuj01} A. Fujiwara, Phys. Rev. A {\bf 63}, 042304 (2001);
A. Fujiwara, H. Imai, J. Phys. A {\bf 36}, 8093 (2003).
\bibitem{Zhe06} J. Zhenfeng et al., preprint LANL quant-ph/0610060
\bibitem{Mon07} {A. Monras, M. G. A. Paris} Phys. Rev. Lett. {\bf 98}, 160401 (2007).
\bibitem{mars} P.~Marian, Phys. Rev. A {\bf 45} 2044 (1992);
P.~Marian, and T.~A.~Marian, Phys. Rev. A {\bf 47}, 4474 (1993);
{\em ibid.} 4487 (1993).
\bibitem{Mab01} J. M. Geremia et al., Phys. Rev. Lett. {\bf 91}, 250801 (2001).
\bibitem{Cra46} H. Cramer, {\em Mathematical methods of statistics},
(Princeton University Press, 1946).
\bibitem{phfed} G. M. D'Ariano, M. G. A. Paris, R. Seno,
Phys. Rev. A  {\bf 54}, 4495 (1996).
\end{thebibliography}
\end{document}